\documentclass{osa-article}

\journal{oe}
\articletype{Research Article}
\begin{document}

\title{Single-frequency mid-infrared chip laser \\ with a sub-hertz Lorentzian linewidth}

\author{Philippe Guay,\authormark{1,2,*} J\'{e}r\^{o}me Genest,\authormark{1} \\ Vincent Michaud-Belleau,\authormark{1} Nicolas Bourbeau H\'{e}bert,\authormark{1} \\ and David G. Lancaster\authormark{2}}

\address{\authormark{1}Centre d'optique, photonique et laser, Universit\'{e} Laval, Qu\'{e}bec, Qu\'{e}bec G1V 0A6, Canada\\
\authormark{2}Laser Physics and Photonics Devices Laboratory, Future Industries Institute, and School of Engineering, University of South Australia, Mawson Lakes, SA 5095, Australia\\
}

\email{\authormark{*}philippe.guay.4@ulaval.ca} %% email address is required



\begin{abstract}
A guided-wave chip laser operating in a single longitudinal mode at 2860 nm is presented. The cavity was set in the Littman-Metcalf configuration to achieve single-frequency operation with a side-mode suppression ratio above 33 dB. The chip laser's linewidth was found to be limited by mechanical fluctuations, but its Lorentzian contribution was estimated to be lower than 1 Hz using a heterodyne technique. This demonstration incorporates a high coherence source with the simplicity provided by the compactness of chip lasers. 
\end{abstract}

%%%%%%%%%%%%%%%%%%%%%%%%%%  body  %%%%%%%%%%%%%%%%%%%%%%%%%%
\section{Introduction}

Single-frequency (SF) lasers, which are sources operating in a single longitudinal mode (SLM), offer low intensity noise and long coherence properties granted by their narrow linewidth. These valued attributes have enabled the use of SF lasers in a broad range of applications such as atomic physics \cite{BAM87,MIC09}, spectroscopy \cite{WYS05,RIC00}, sensing \cite{HEN17}, LIDAR \cite{DRA02}, and gravitational wave detection \cite{QI18}. Many of these applications, namely spectroscopy and sensing, would greatly benefit from laser sources in the mid-infrared (MIR), a spectral region where generation of coherent light has always been challenging. 

Many techniques have been developed to achieve single-mode operation in the MIR. Optical parametric oscillators (OPOs) have been extensively used \cite{LAI18,SHU18,WAN16,ZHA18}. They offer wide wavelength tuning range, high output power, narrow linewidth, and high-stability, but they use a nonlinear crystal and a cavity in which an etalon is inserted to suppress undesired modes, which quickly escalates the level of complexity. As a result, OPO-based SF lasers often become cumbersome and expensive instruments. 

Generation of light in the MIR region can also be achieved through difference frequency generation (DFG). DFG lasers are well established in the  3 - 4.5 $\mu$m range \cite{LAN99,INS16,VAS08} and have the potential for a complete coverage of the mid-infrared region. However, the schemes used to produce MIR radiation are bulky as DFG often requires multiple sources of radiation and amplifiers. Additionally, the low output power reached so far with SF lasers remains an issue for applications with high power requirement such as atmospheric sensing.
 
Quantum cascade lasers (QCLs) stand as a reliable and compact alternative to produce MIR radiation in the 3-25 $\mu$m region. To reach single-frequency operation, the lasers are often used with an external cavity \cite{WAN17,TOT02} based on Littrow configuration. Alternatively, QCLs are used with an integrated reflective Bragg grating \cite{YAO12} as a wavelength selective feedback element. In any cases, the internal dynamics of semiconductor lasers is fast compared to that of solid-state lasers. This can leads to significant phase noise that, in addition to the coupling between intensity and phase noise \cite{HEN82}, explains the broad linewidth of semiconductor lasers. Typically, the linewidth of a semiconductor laser is in the order of MHz \cite{CAL07} while the linewidth of solid-state lasers is in the order of a few kHz \cite{YAN19}.

Many of the challenges mentioned above are addressed by fiber lasers. They provide simplicity, flexibility, adequate heat management, and excellent output beam quality. Most demonstrations of SF fiber lasers have been conducted in the near-infrared up to 2 $\mu$m \cite{HEN17,SHI19,WU09,LIM16,QI17}. So far, only a few have reached laser emission in the 3 $\mu$m region \cite{BER15,HUD13}. Similar performance have been reached with MIR ring cavity lasers \cite{COL12} based on Cr:ZnSe, but there has not been any demonstrations of SF emission with a Lorentzian linewidth below 20~kHz.  

In this paper, we present a Ho$^{3+}$ and Pr$^{3+}$-codoped ZBLAN chip laser \cite{LAN19} that is operated in the single-frequency regime with  a Lorentzian linewidth measured to be below 1 Hz. This laser embodies the qualities of chip-based instruments such as simplicity and compactness. The laser also overcomes any size limitation encountered in fiber lasers due to the bending radius of the fluorozirconate fiber or overcomes the difficulty of handling a small length of fiber.  As demonstrated in \cite{NBH18}, chip lasers are excellent candidates to become compact instruments. There has been a demonstration of a MIR SF compact waveguide laser \cite{YOU15}, but there has yet to be a measurement of the Lorentzian linewidth of the single longitudinal mode. In addition to the linewidth measurement provided in this paper, the single-mode operation of the laser is confirmed by the analysis of the laser's spectral emission with a scanning Fabry-Perot interferometer (FPI). The FPI is also used to provide an approximation of the side-mode suppression ratio (SMSR), which is validated by the mixing of the laser's longitudinal mode. 

\section{Experimental setup}\label{exp_set}

\begin{figure}[h]
\centering
\includegraphics[width=0.8\linewidth]{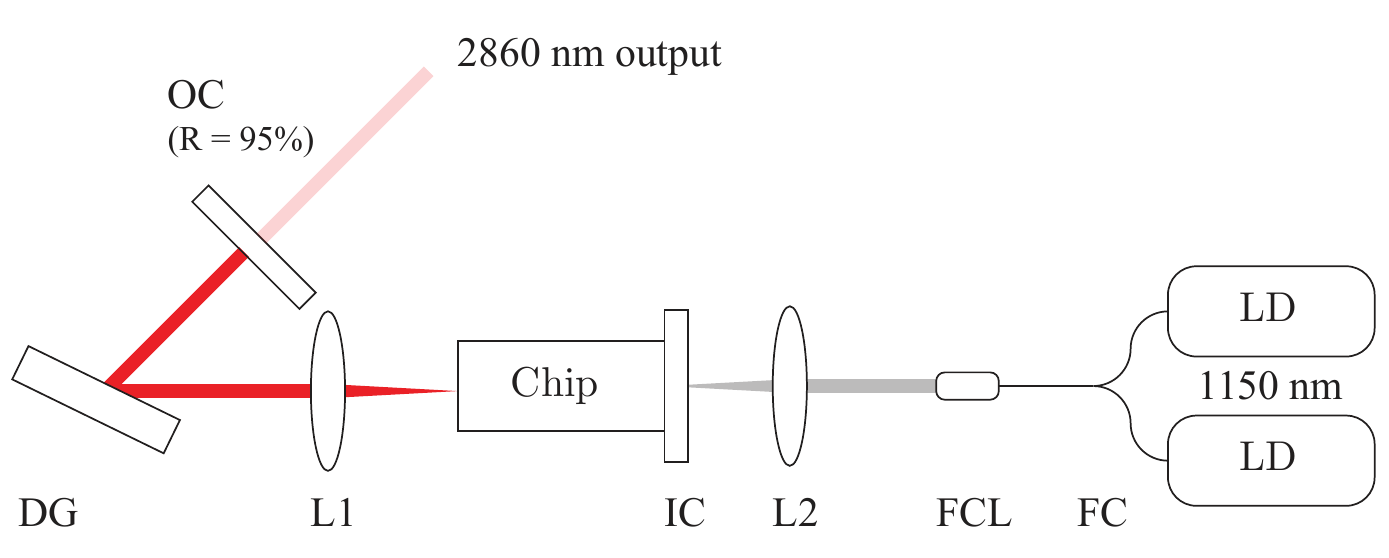}
\caption{Experimental setup of the Littman-Metcalf cavity. DG: Diffraction grating. OC: 95\% output coupler.  L1: Lens ($f~=~50$ mm). IC: Input coupler. L2: Lens ($f$~=~50 mm). FCL: Fiber-coupled lens ($f$~=~6.51 mm). FC: Fiber combiner. LD: 1150 nm laser diode.}
\label{fig:exp_set}
\end{figure}

The experimental setup is shown in Fig.~\ref{fig:exp_set}. The laser is based on a 11 mm holmium-active fluorozirconate glass chip producing optical radiation at 2.86 $\mu$m when pumped at 1150 nm. The two pump diodes (Innolume LD-1152-FBG-370), which output a total power of 800 mW, are polarization-combined, collimated, and focused on the chip's depressed cladding waveguide \cite{LAN13}. The lenses prior to the chip are chosen to ensure adequate mode matching between the fiber and the 100 $\mu$m diameter waveguide. The beam coming out of the chip is collimated to illuminate a grating (Thorlabs GR1325-45031) with 450 lines/mm, a blaze angle of 32$^\text{o}$, and a dispersion of 1.6 nm/mrad, positioned to reflect the first diffraction order onto a 95\% reflective output coupler. The latter reflects the signal back to the grating and transmits 12 mW of 2.9~$\mu$m light. A slope efficiency of nearly 3\%  is achieved.

As the first order of diffraction is reflected toward an output coupler rather than back at the cavity as it would in the Littrow configuration, the angular dispersion of the retroreflected light is twice that of the light reflected in a single pass. Such an arrangement, that is the Littman-Metcalf configuration, improves the side-mode suppression ratio (SMSR) and helps reaching single-mode operation.

%\begin{figure}[h!]
%\centering\includegraphics[width=7cm]{osafig1}
%\caption{Sample caption (Fig. 2, \cite{Yelin:03}).}
%\end{figure}

\section{Side-mode suppression ratio}

% intro to the 2 techniques
Single-frequency operation of the chip laser was confirmed by observation of its spectrum using a scanning Fabry-Perot interferometer (FPI) and validated by the analysis of the beatnote arising from the mixing of the main longitudinal mode with a strongly attenuated side-mode.

\begin{figure}[h]
\centering
\includegraphics[width=0.8\linewidth]{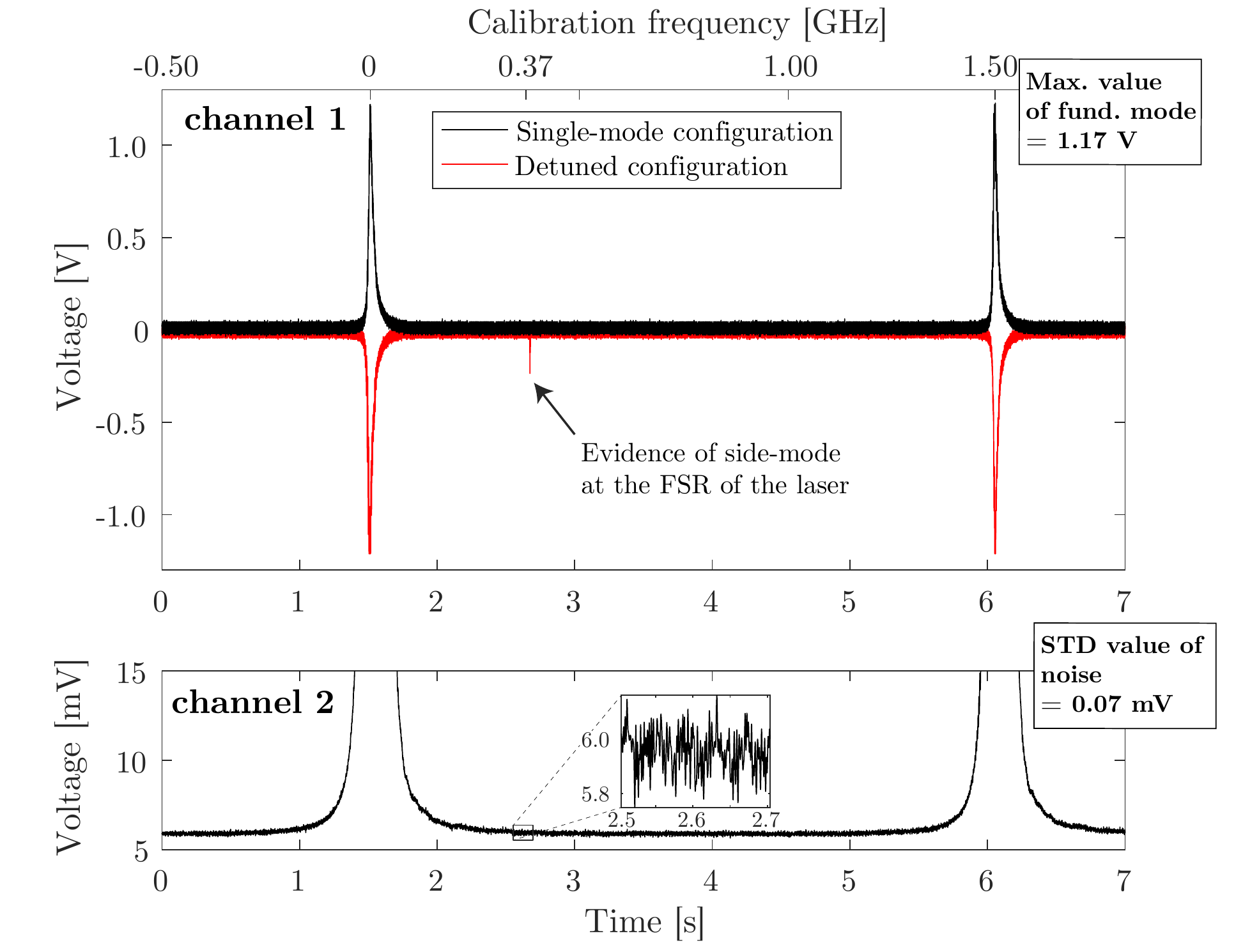}
\caption{(top) Scanned spectrum obtained using the FPI with a scaling adapted to measure the peak amplitude of the fundamental mode for single-mode (black) and detuned (red) configurations. (bottom) Scanned spectrum on a scaling adapted to measure a low intensity signal and zoomed-in (inset) portion where the side-mode was previously observed.}
\label{fig:FP}
\end{figure}

\subsection{Fabry-Perot interferometry}

% FPI description
A scanning FPI utilized as a spectrum analyzer is a useful tool to observe a laser's fine spectral features. As such, this instrument (Thorlabs SA200-30C) with a finesse of 290 and a free spectral range of 1.5 GHz has been used here to examine the longitudinal modes of the chip laser set in two different configurations. In the first configuration, the grating in the cavity was purposely detuned to reveal the presence of a side-mode approximately 371 MHz away from the main mode, in agreement with the expected free spectral range of the 40-cm-long chip laser cavity. The result of this measurement is plotted, as negative for comparaison purposes, in Fig.\ref{fig:FP}. The main mode displays a linewidth of 11 MHz, which is limited by the finesse  of the FP cavity. The slight asymmetry in the fundamental mode's lineshape is attributed to a voluntary misalignment of the FPI cavity to remove undesired reflections back into the chip laser. No optical isolator was available to eliminate stray reflections back into the chip laser. Another measurement shown in black in Fig.~\ref{fig:FP}~(a) was made in a finely tuned configuration where no side-mode could be seen. The grating was rotated using a stepper motor to spectrally overlap the center frequency of the grating with the fundamental mode. The trace in this configuration was examined on two oscilloscope channels to simultaneously measure the peak value of the main mode and the low intensity noise at the location where the side-mode was previously observed. The main mode's maximum amplitude measured at 1.17 V and the smallest signal detectable in the noise allows an estimation of the SMSR. The standard deviation of the noise estimated at 0.07 mV is used as an indicator of the highest amplitude of a low intensity signal that could be detected, thus a SMSR ratio of at least 42 dB was obtained.

% Additionnal information : observation of mode-hop
While studied with the scanning FPI, the laser has been observed to mode-hop on a scale of a few seconds. Mechanical and acoustic vibrations as well drifts of the chip's temperature could explain the instability in the wavelength selectivity of the grating. Moreover, many optical components were mounted on long optical posts. It has been shown in \cite{NBH18} for a similar chip laser that mechanical integration of the laser has greatly reduced the impact of environmental fluctuations.

\subsection{Longitudinal modes beatnote}

% intro to the other technique
The SMSR ratio obtained via Fabry-Perot interferometry was validated by analyzing the beat of the longitudinal modes of the chip laser. The laser's output was sent to a MCT photodetector with 350 MHz bandwidth (PEMI-10.6) to observe interference between modes. As a result, the relative amplitude of the modes was extracted and the SMSR was calculated. Since only one side-mode was observed on the FPI scan for a detuned cavity, the following analysis assumes the presence a of single side-mode. 

\begin{figure}[h]
\centering
\includegraphics[width=0.9\linewidth]{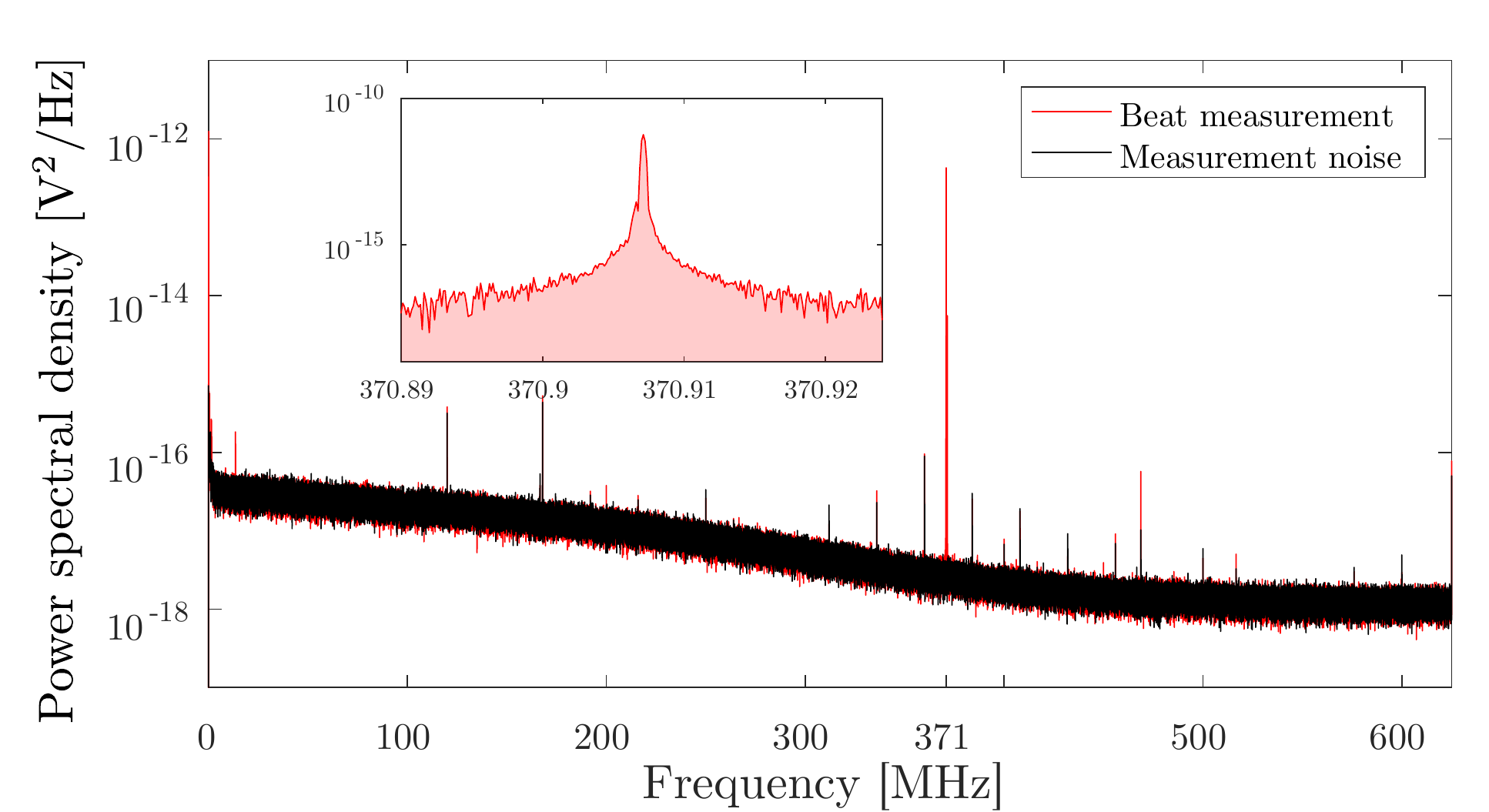}
\caption{Voltage power spectral density of the chip laser at the detector's output in the absence (black curve) and presence (red curve) of laser light where a beat note is visible at the expected 371 MHz FSR of the 40-cm-long cavity. (inset) Zoom-in on the beatnote to highlight the red area representing the power in the mixing term.}
\label{fig:beatnote}
\end{figure}

% description of the technique + results
Let the strongest mode be represented by an electric field $E_1\cos{(\omega_1 t)}$ and its closest secondary mode by $E_2\cos{(\omega_2 t)}$. The sum of these fields is squared by the measurement of the resulting field intensity. Within the bandwidth of the photodetector, the measured voltage on the oscilloscope is therefore proportional to $E_1^2 + E_2^2 + 2E_1E_2\cos{((\omega_2-\omega_1) t)}$, where the first two terms represent the DC value and the last represents the mixing term at a frequency which corresponds to the FSR of the chip laser's cavity length.  The integrated power in the tone located at 371 MHz (shaded area in inset of Fig. \ref{fig:beatnote}) normalized by the square of the DC value obtained through a time-average of the waveform provides the ratio $(E_2/E_1)^2$, which equals the SMSR in our simple model. The SMSR is estimated here at 33 dB. 

% Comparison between FPI and beatnote

The difference-frequency signal $2E_1E_2\cos{((\omega_2-\omega_1) t)}$, which is used to estimate the amplitude of the low intensity side-mode, results from the product of the high amplitude main mode $E_1$ and the low amplitude side-mode $E_2$. Thus, the difference-frequency term has a larger amplitude than the side-mode, which allows a better estimation of the low intensity signal. However, the diffraction grating was not in the same experimental conditions as with the FPI measurement : the angular position of the grating was different and the measurement was done on a different day. Also, the presence of an another attenuated side-mode unaccounted for in the model could slightly impact the results. Consequently, it is expected that the ratio found here be different than with the FPI scan. Nevertheless, the ratio obtained for a cavity tuned without direct insight on the side-mode amplitude is to be treated as a minimum value that confirms the FPI result.

\section{Linewidth measurement}

\begin{figure}
\centering
\includegraphics[width=0.9\linewidth]{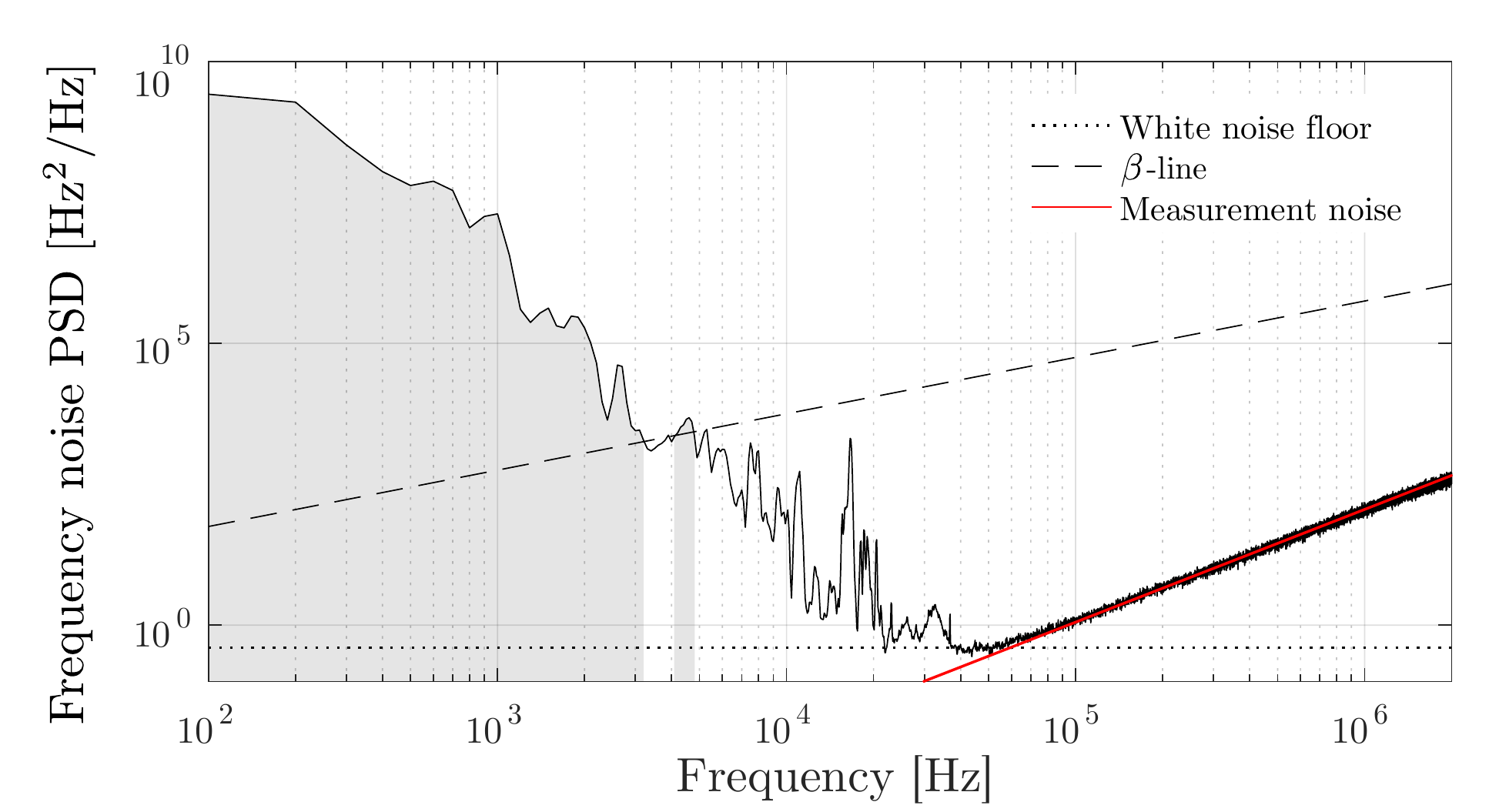}
\caption{Frequency noise PSD of the beat note from the Littman-Metcalf and Littrow lasers. The grey area is used to determine the linewidth over the total measurement duration while the noise level around 40 kHz provides an upper bound to the Lorentzian linewidth.}
\label{fig:psd_fre_noi}
\end{figure}

An upper limit to the laser's linewidth was estimated from the heterodyne beating of the SF laser studied here with a similar but Littrow-configured, and thus multi-longitudinal mode laser \cite{LAN19}. Both lasers have been directed toward a free-space beam combiner and sent to a TE cooled MCT photodetector with 17.5 MHz bandwidth (PVI-2TE4). The beams were carefully aligned to ensure a reasonable spatial overlap between them. The chip laser's frequency was finely tuned by rotating the grating to approach one of the few modes of the Littrow laser having a FSR of 3 GHz. The frequency noise power spectral density (PSD) of the beat note shown in Fig.~\ref{fig:psd_fre_noi} can provide an upper bound to the laser's linewidth, but one should note that this measurement sums the frequency noise of both lasers and therefore only yields the desired spectrum if the reference laser is significantly more stable, which cannot be assumed here. However, since the multi-mode laser was built using similar chip and diffraction grating than those in the SF laser and that the single-mode laser and since both lasers were operated in the same gain condition, it is expected that their modes show a similar linewidth. In any event, the measurement given here provides an upper bound for the SF laser linewidth.

The laser's linewidth can be obtained from reference~\cite{DOM10}, which estimates it as $(8\ln{(2)}A)^{1/2} = 2~\text{MHz}$ over a measurement time of 10 ms, where $A$ is the area under the portions of the frequency noise PSD that exceed the $\beta$-line. This linewidth is dominated by frequency fluctuations below 5 kHz. These come from mechanical and acoustic vibrations exacerbated by the incomplete mechanical integration of the laser. It is expected that integration such as in \cite{NBH18} will greatly reduce the impact of these environmental fluctuations and help reach the white frequency noise level over a broader frequency range. Sharp spurs observed between 7 kHz and 40 kHz on the frequency noise PSD have only been observed for measurements where the grating's angular position was controlled with a stepper motor. It can therefore be concluded that these arise from mechanical resonances of the actuators. The rising $f^2$ slope at frequencies over 20 kHz stems from additive measurement noise converted to the frequency noise PSD, its level being in agreement with the model of additive noise converted to frequency noise, shown in red. The Lorentzian linewidth corresponding to the intrinsic linewidth of the fundamental mode can also be estimated from the frequency noise PSD. Its full width half maximum (FWHM) is given as FWHM~=~$\pi h_0=$~1~Hz, where $h_0$ is given as the white noise floor on the frequency noise PSD. Here, the white noise floor is barely distinguishable as the curve reaches a minimum value where the mechanical fluctuations meet the measurement noise near 50 kHz. Nevertheless, this section of the black curve can be used to determine an upper bound for the Lorentzian linewidth. 

In conclusion, the operation of a single-frequency laser based on a chip at 2.9 $\mu$m was demonstrated. The laser was characterized in terms of its side-mode suppression ratio and linewidth. The Lorentzian linewidth was found to be below the Hz level, which is less than the linewidth of any other mid-infrared sources reported so far.

\section*{Funding}
Natural Sciences and Engineering Research Council of Canada (NSERC) ; Fonds de Recherche du Qu\'{e}bec-Nature et Technologies (FRQNT); The University of South Australia team gratefully acknowledges start up funding support form the University. 
\\

\section*{Disclosures}

The authors declare no conflicts of interest.

\end{document}